\documentstyle[leqno,twoside,11pt,proc209]{article} 
\begin{document}
\cleardoublepage
\thispagestyle{empty}
\pagestyle{empty}
\title{Invariants and Symmetries for Partial Differential Equations and 
Lattices\thanks{Research supported in part by NSF 
under Grant CCR-9625421.}}
\author{\"Unal G\"okta\c{s}\thanks{Colorado School of Mines,
Dept. of Mathematical and Computer Sciences, Golden, CO 80401-1887}
\and 
Willy Hereman$^\dagger$}
\date{}
\maketitle
\thispagestyle{empty}
\pagestyle{empty}
\begin{abstract}
Methods for the computation of invariants and symmetries of nonlinear 
evolution, wave, and lattice equations are presented. 
The algorithms are based on dimensional analysis, and can be 
implemented in any symbolic language, such as {\it Mathematica}.  
Invariants and symmetries are shown for several well-known equations.

Our {\it Mathematica\/} package allows one to automatically compute 
invariants and symmetries. Applied to systems with parameters, 
the package determines the conditions on these parameters so that a 
sequence of invariants or symmetries exists. 
The software can thus be used to test the integrability of 
model equations for wave phenomena. 
\end{abstract}

\section{The Key Concept: Scaling Invariance}

The ubiquitous Korteweg-de Vries (KdV) equation from soliton theory, 
\begin{equation}
\label{kdv}
u_t = 6 u u_x + u_{3x},
\end{equation} 
is invariant under the dilation (scaling) symmetry
$ (t, x, u) \rightarrow ({\lambda}^{-3} t, \lambda^{-1} x, {\lambda}^{2} u),$
where $\lambda$ is an arbitrary parameter. 
Obviously, $u$ corresponds to two derivatives in $x$, 
i.e.\ $u \sim {\partial^2}/{\partial {x^2}}.$
Introducing weights, $w(u) = 2$ if we set $w({\partial}/{\partial x}) = 1.$ 
Similarly, $ {\partial}/{\partial t} \sim {\partial}^3/{\partial x}^3,$ 
thus $w({\partial}/{\partial t})=3.$ 
The {\it rank} of a monomial equals the sum of all of its weights. 
Observe that (\ref{kdv}) is {\it uniform in rank} since all the terms
have rank $R=5$. 

Likewise, the Volterra lattice, which is one of the discretizations of 
(\ref{kdv}),
\begin{equation} \label{volterra}
{\dot{u}}_n = u_n \, (u_{n+1}-u_{n-1}), 
\end{equation}
is invariant under $(t, u_n) \rightarrow (\lambda^{-1} t, \lambda u_n).$
So, $u_n \sim {\rm d}/{\rm dt},$ or $w(u_n)=1$ if we set 
$w({\rm d}/{\rm dt})=1.$ Every term in (\ref{volterra}) has rank $R=2,$ 
thus (\ref{volterra}) is uniform in rank.

Scaling invariance, which is a special Lie-point symmetry, is common to 
many integrable nonlinear partial differential equations (PDEs) 
such as (\ref{kdv}), and nonlinear differential-difference equations (DDEs) 
like (\ref{volterra}).
Both equations have infinitely many polynomial invariants
\cite{ghjsc,ghepla} and symmetries \cite{baltzer}.
In this paper we show how to use the scaling invariance to explicitly 
compute polynomial invariants and symmetries of PDEs and DDEs. 

\section{Computation of Invariants}

For PDE systems as in Table 1, the conservation law 
${\rm D}_{t} \rho = {\rm D}_{x} J $ connects the {\em invariant} 
({\rm conserved density\/}) $\rho$ and the associated {\em flux\/} $J.$ 
As usual, ${\rm D}_{t}$ and ${\rm D}_{x}$ are total derivatives.
Most {\it polynomial} density-flux pairs only depend on ${\bf u}, 
{\bf u}_x, $ etc. (not explicitly on $t$ and $x).$
\samepage{
Integration of the conservation law with respect to $x$ yields that
$ P = \int_{-\infty}^{+\infty} \rho \; dx $ is constant in time, 
provided $J$ vanishes at infinity. $P$ is a conserved quantity.

The first three (of infinitely many) conservation laws for (\ref{kdv}) are 
\begin{eqnarray} \label{kdvconslaw1and2}
& & {\rm D}_t (u)  = {\rm D}_x (3 u^2 + u_{2x} ),\quad 
{\rm D}_t (u^2) = {\rm D}_x ( 4 u^3 - u_x^2 + 2 u u_{2x} ), \\
\label{kdvconslaw3}
& & {\rm D}_t (u^3 - \frac{1}{2} u_x^2) =
{\rm D}_x ({9 \over 2} u^4 - 6 u u_x^2 + 3 u^2 u_{2x} 
+ \frac{1}{2} u_{2x}^2 - u_x u_{3x}).
\end{eqnarray}
The densities $\rho = u, u^2, u^3 - \frac{1}{2} u_x^2 $ have 
ranks $2, 4$ and $6$, respectively.
\vskip 3pt
\noindent
Conserved densities of PDEs like (\ref{kdv}) can be computed as follows:
\vskip 3pt
\noindent
$\bullet$ Require that each equation in the PDE system is uniform in rank.
Solve the resulting linear system to determine 
the weights of the dependent variables. For (\ref{kdv}), solve 
$ w(u) + w({\partial}/{\partial t}) = 2 w(u) + 1 = w(u) + 3, $ to get 
$ w(u) = 2$ and $w({\partial}/{\partial t}) = 3.$ 
\vskip 3pt
\noindent
$\bullet$
Select the rank $R$ of $\rho ,$ say, $R=6.$ 
Make a linear combination of all the monomials in the components of 
${\bf u}$ and their $x$-derivatives that have rank $R.$ 
Remove `equivalent' monomials, that is, those that are total $x$-derivatives
(like $u_{4x}$) or differ by a total $x$-derivative. 
For example, $u u_{2x} $ and $u_x^2$ are equivalent 
since $u u_{2x} = \frac{1}{2} (u^2)_{2x} - u_x^2.$ 
For (\ref{kdv}), one gets $\rho = c_1 u^3 + c_2 u_x^2 $ of rank $R=6.$
\vskip 3pt
\noindent
$\bullet$
Substitute $\rho$ into the conservation law, eliminate all $t$-derivatives, 
and require that the resulting expression is a total $x$-derivative. 
Apply the Euler operator \cite{ghjsc} to avoid integration 
by parts. The remaining part must vanish identically. 
This yields a linear system for the constants $c_i.$ Solve the system. 
For (\ref{kdv}), one gets $c_1 = 1, c_2 = -1/2.$
\vskip 3pt
\noindent
See \cite{ghjsc} for the complete algorithm and its implementation.
See \cite{soft} for an integrated {\em Mathematica}
Package that computes invariants (and also symmetries) of PDEs and DDEs. 
\begin{table}[hp] \caption{{\rm Invariants and Symmetries}}
\vskip .001pt
\noindent
\begin{tabular}{ | l | l | l |} \hline 
& Continuous Case (PDEs) & Semi-discrete Case (DDEs) \\ [0.5ex] 
\hline
& & \\
System & 
${\bf u}_t={\bf F}({\bf u}, {\bf u}_{x}, {\bf u}_{2x}, ...)$ &
${\dot{\bf u}}_n={\bf F}(...,{\bf u}_{n-1}, {\bf u}_{n}, {\bf u}_{n+1},...)$
\\
& & \\
\hline 
& & \\
Cons. Law & $ {\rm D}_{t} \rho = {\rm D}_{x} J$ &
${\dot{\rho}}_n=J_n - J_{n+1}$ \\
& & \\
\hline 
& & \\
Symmetry & ${\rm D}_t{\bf G} \!=\! {\bf F}'({\bf u})[{\bf G}] 
\!=\!{\partial\over\partial{\epsilon}}{\bf F}({\bf u}
+\epsilon {\bf G})|_{\epsilon=0} $ & 
$ {\rm D}_t{\bf G}\!=\!{\bf F}'({\bf u}_n)[{\bf G}]\!=\! {\partial 
\over \partial{\epsilon}}{\bf F}({\bf u}_n 
+\epsilon {\bf G})|_{\epsilon=0}$\\
& & \\
\hline 
\end{tabular}
\end{table}
\vskip .01pt
\noindent
For DDEs like (\ref{volterra}), compute the weights in a similar way. 
Determine all monomials of rank $R$ in the components of ${\bf u}_n$ and 
their $t$-derivatives. Use the DDE to replace all the $t$-derivatives.
Monomials are `equivalent' if they belong to the same equivalence class
of shifted monomials. Keep only the main representatives (centered at $n$) 
of the various classes.
Combine these representatives linearly with coefficients $c_i,$
and substitute the form of $\rho_n$ into the conservation law 
$\dot \rho_n = J_n - J_{n+1}. $
Remove all $t$-derivatives and pattern-match the resulting expression
with $J_n - J_{n+1}.$ 
Set the non-matching part equal to zero, and solve the linear system 
for the $c_i.$ Determine $J_n$ from the pattern $J_n - J_{n+1}.$ 
For (\ref{volterra}), the first three (of infinitely many) densities 
$\rho_n$ are listed in Table 2. 
Details about the algorithm and its implementation are 
in \cite{physicad,soft,ghepla}. 
} 
\vfill
\newpage
\section{Computation of Symmetries}

As summarized in Table 1, 
${\bf G} (x, t, {\bf u}, {\bf u}_{x}, {\bf u}_{2x}, ...)$
is a {\it symmetry\/} of a PDE system iff it leaves it invariant for 
the change ${\bf u} \rightarrow {\bf u} + \epsilon {\bf G}$ within order 
$\epsilon.$ Hence, 
$ {\rm D}_t ({\bf u} + \epsilon {\bf G}) = 
{\bf F} ({\bf u} + \epsilon {\bf G}) $
must hold up to order $\epsilon.$ 
Thus, ${\bf G}$ must satisfy the linearized equation 
$ {\rm D}_t {\bf G} = {\bf F}'({\bf u})[{\bf G}], $ 
where ${\bf F}'$ is the Fr\'echet derivative:
${\bf F}'({\bf u})[{\bf G}] = {\partial \over \partial{\epsilon}} 
{\bf F}({\bf u}+\epsilon {\bf G}) |_{\epsilon = 0}.$

Using the dilation invariance, generalized symmetries ${\bf G}$ can 
be computed as follows:
\vskip 3pt
\noindent
$\bullet$ Determine the weights of the dependent variables as in Section 2.
\vskip 3pt
\noindent
$\bullet$ Select the rank $R$ of the symmetry. 
Make a linear combination of all the monomials involving ${\bf u}$ and
its $x$-derivatives of rank $R.$ 
For example, for (\ref{kdv}), 
$G = c_1 \, u^2 u_x + c_2 \, u_x u_{2x} + c_3 \, u u_{3x} + c_4 \, u_{5x}$
is the generalized symmetry of rank $R=7.$
\vskip 3pt
\noindent
$\bullet$ Compute ${\rm D}_t {\bf G}.$ 
Use the PDE system to remove all $t$-derivatives. 
Equate the result to the Fr\'echet derivative ${\bf F}'({\bf u})[{\bf G}].$
Treat the different monomial terms in ${\bf u}$ and its $x$-derivatives 
as independent, to get the linear system for $c_i.$ Solve that system. 
For (\ref{kdv}), one obtains
\begin{equation}
\label{laxsymm}
G =  30 u^2 u_x + 20 u_x u_{2x} + 10 u u_{3x} + u_{5x}. 
\end{equation}

The symmetries of rank 3, 5, and 7 are listed in Table 2. They are the 
first three of infinitely many. 
\begin{table}[hp] \caption{{\rm Prototypical Examples}}
\vskip 2pt
\noindent
\begin{tabular}{ | l | l | l |} \hline 
& Korteweg-de Vries Equation & Volterra Lattice \\ [0.5ex] 
\hline
& & \\
Equation & 
$u_t = 6 u u_x + u_{3x}$ &
${\dot{u}}_n = u_n \, (u_{n+1}-u_{n-1})$\\
& & \\
\hline 
& & \\
Invariants & $\rho\!=\!u \quad\quad\;\;\;\;\;\;\;\quad \rho=u^2 $&
${\rho}_n=u_n \quad\quad\;\;{\rho}_n=u_n (\frac{1}{2}u_n + u_{n+1})$\\
& & \\
&$\rho=u^3-\frac{1}{2} u_x^2 $&
${\rho}_n\!=\!\frac{1}{3}u_n^3\!+\!u_n u_{n+1}(u_n\!+\!u_{n+1}\!+\!u_{n+2})$ \\
& & \\
\hline 
& & \\
Symmetries & ${\bf G}=u_x \quad\;\;\;\;\;\quad\;\;\;\; 
{\bf G}=6 u u_x + u_{3x}$ &
${\bf G}=u_n u_{n+1} (u_n + u_{n+1} + u_{n+2})$\\
&${\bf G}\!\!=\!30 u^2u_x\!+\!20u_xu_{2x}\!+\!10 uu_{3x}\!+\!u_{5x}$ & 
$\;\;\;\;\;\;\;\;- u_{n-1} u_n (u_{n-2} + u_{n-1} + u_n) $\\
& & \\
\hline 
\end{tabular}
\end{table}
\noindent
For DDEs like (\ref{volterra}), 
${\bf G} (...,{\bf u}_{n-1}, {\bf u}_{n}, {\bf u}_{n+1},...) $ is a 
{\it symmetry\/} iff the infinitesimal transformation
$ {\bf u}_n \rightarrow {\bf u}_n + \epsilon 
{\bf G}$ leaves the DDE invariant within order $\epsilon$. 
Consequently, ${\bf G}$ must satisfy
$ {{\rm d}{\bf G} \over {\rm d}{t}} = {\bf F}'({\bf u}_n)[{\bf G}], $
where ${\bf F}'$ is the Fr\'echet derivative,
$ {\bf F}'({\bf u}_n)[{\bf G}] = {\partial \over \partial{\epsilon}} 
{\bf F}({\bf u}_n +\epsilon {\bf G}) |_{\epsilon = 0}. $

Algorithmically, one performs the following steps: First compute the weights
of the variables in the DDE. 
Determine all monomials of rank $R$ in the components of ${\bf u}_n$ and 
their $t$-derivatives. Use the DDE to replace all the $t$-derivatives.
Make a linear combination of the resulting monomials with coefficients $c_i.$
Compute $D_t {\bf G}$ and remove all $ {\dot {\bf u}}_{n-1}, 
{\dot {\bf u}}_n, {\dot {\bf u}}_{n+1},$ etc. 
Equate the resulting expression to the Fr\'echet derivative 
$ {\bf F}'({\bf u}_n)[{\bf G}] $ and solve the system for the $c_i,$
treating the monomials in ${\bf u}_n$ and its shifts as independent. 
Details are in \cite{physicad,ghepla}. 
For (\ref{volterra}), the symmetry ${\bf G}$ of rank $R=3$ is listed 
in Table 2. There are infinitely many symmetries, all with different ranks.

See \cite{baltzer} for the complete algorithm and its implementation in 
{\it Mathematica}, 
and \cite{soft} for an integrated {\em Mathematica} Package that
computes symmetries of PDEs and DDEs.
\vskip 3pt
\noindent
{\bf Notes:} 
\vskip 2pt
\noindent
(i) If PDEs or DDEs are of second or higher order in $t$, 
like the Boussinesq equation in \cite{ghjsc}, we assume that they
can be recast in the form given in Table 1. 
\vskip 2pt
\noindent
(ii) A slight modification of the methods in Section 2 and 3 allows one to 
find invariants and symmetries that explicitly depend on $t$ and $x.$
See next section for an example. 
\vskip 2pt
\noindent
(iii) Applied to systems with free parameters, the linear system for the 
$c_i$ will depend on these parameters. A careful analysis of the eliminant 
leads to conditions on these parameters so that a sequence of invariants 
or symmetries exists. 
\vskip 2pt
\noindent
(iv) For equations that lack uniformity in rank, we introduce one or more
auxiliary (constant) parameters with weights. 
After the form of the invariant or symmetry is determined, 
the auxiliary parameters can be reset to one. 
\vskip 2pt
\noindent
(v) Higher-order symmetries, such as (\ref{laxsymm}) lead to new 
integrable evolution equations. For example, 
$ u_t =  30 u^2 u_x + 20 u_x u_{2x} + 10 u u_{3x} + u_{5x}$ 
is the completely integrable fifth-order KdV equation due to Lax. 
\vskip 3pt
\noindent
Details about these 5 items are given in \cite{ghjsc,physicad,baltzer,ghepla}.

\section{Examples}

\subsection{Vector Modified KdV Equation}

In \cite[Eq.\ (4)]{verheest}, Verheest investigates the integrability of
a vector form of the modified KdV equation (vmKdV), which upon projection, 
reads
\begin{eqnarray}
\label{vmkdv}
u_t + 3 u^2 u_x + v^2 u_x + 2 u v v_x + u_{3x}  = 0, 
\nonumber \\
v_t + 3 v^2 v_x + u^2 v_x + 2 u v u_x + v_{3x} = 0.
\end{eqnarray}
With our software {\bf InvariantsSymmetries.m} \cite{soft} 
we computed the following invariants:
\begin{eqnarray}
\rho_1 &=& u, \quad \rho_2 = v, \quad \rho_3 = u^2 + v^2, \\
\rho_4 &=& \frac{1}{2} (u^2 + v^2)^2 - (u_x^2 +  v_x^2),  \\
\rho_5 &=& 
\frac{1}{3} x (u^2 + v^2) - \frac{1}{2} t (u^2 + v^2)^2 + t (u_x^2 + v_x^2).
\end{eqnarray}
Note that the latter invariant depends explicitly on $x$ and $t.$
Verheest \cite{verheest} has shown that (\ref{vmkdv}) is non-integrable
for it lacks a bi-Hamiltonian structure and recursion operator. 
We were unable to find additional polynomial invariants. 
Polynomial higher-order symmetries for (\ref{vmkdv}) do not appear to exist. 

\subsection{Extended Lotka-Volterra Equation}

Itoh \cite{itoh} studied the following extended version of 
the Lotka-Volterra equation (\ref{volterra}), 
\begin{equation}\label{extvolterra}
{\dot{u}}_n = \sum_{r=1}^{k-1} (u_{n-r} - u_{n+r}) u_n .
\end{equation}
For $k=2,$ (\ref{extvolterra}) is (\ref{volterra}), for which three
invariants and one symmetry are listed in Table 2. 
In \cite{ghepla}, we give two additional invariants;
in \cite{baltzer} we list two more symmetries. 
\vskip 2pt
\noindent
For (\ref{extvolterra}), we computed 5 invariants and 2 higher-order 
symmetries for $k=3$ through $k=5.$ 
\vfill
\newpage
\noindent
Here is a partial list of our results: 
\vskip 3pt
\noindent
\noindent
{\bf Case 1:} $k = 3$
\vskip 3pt
\noindent
Invariants:
\vskip 2pt
\noindent
\begin{eqnarray}
\;\; \rho_1 &=& u_n, \quad\quad
\;\;\rho_2 = \frac{1}{2} u_n^2 + u_n (u_{n+1} + u_{n+2}), \\
\;\; \rho_3 &=& \frac{1}{3} u_n^3 + u_n^2 (u_{n+1} + u_{n+2})
+ u_n (u_{n+1} + u_{n+2})^2
\nonumber \\ 
&& + u_n (u_{n+1} u_{n+3} + u_{n+2} u_{n+3} + u_{n+2} u_{n+4}).
\end{eqnarray}
\vskip 2pt
\noindent
Higher-order symmetry:
\vskip 2pt
\noindent
\begin{eqnarray}
\;\;\;\;\;\;\;\; G \!&\!=\!&\! 
u_n^2 (u_{n+1} + u_{n+2} - u_{n-2} - u_{n-1})
+ u_n [(u_{n+1} + u_{n+2})^2 - (u_{n-2} + u_{n-1})^2 ]
\nonumber \\
&&
+ u_n [u_{n+1} u_{n+3} + u_{n+2} u_{n+3} + u_{n+2} u_{n+4}
- (u_{n-4} u_{n-2} + u_{n-3} u_{n-2} + u_{n-3} u_{n-1})]. 
\end{eqnarray}
\vskip 2pt
\noindent
{\bf Case 2:} $k = 4$
\vskip 2pt
\noindent
Invariants:
\vskip 2pt
\noindent
\begin{eqnarray}
\;\;\;\;\;\;\;\;\; \rho_1 \!&\!=\!&\! u_n, \quad\quad
\rho_2 = \frac{1}{2} u_n^2 + u_n (u_{n+1} + u_{n+2} + u_{n+3}), \\
\;\;\;\;\;\;\;\; \rho_3 \!&\!=\!&\! \frac{1}{3} u_n^3 + 
u_n^2 (u_{n+1} + u_{n+2} + u_{n+3}) + u_n (u_{n+1} + u_{n+2} + u_{n+3})^2 
\nonumber \\
&& + u_n (u_{n+1} u_{n+4} + u_{n+2} u_{n+4} + u_{n+3} u_{n+4} 
+ u_{n+2} u_{n+5} + u_{n+3} u_{n+5} + u_{n+3} u_{n+6}).
\end{eqnarray}
\vskip 2pt
\noindent
Higher-order symmetry:
\vskip 2pt
\noindent
\begin{eqnarray}
\;\;\; G \!&\!=\!&\! u_n [
u_{n+1} u_{n+4} + u_{n+2} u_{n+4} + u_{n+3} u_{n+4} 
+ u_{n+2} u_{n+5} + u_{n+3} u_{n+5} + u_{n+3} u_{n+6} 
\nonumber \\
&& - (u_{n-6} u_{n-3} + u_{n-5} u_{n-3} + u_{n-4} u_{n-3}
+ u_{n-5} u_{n-2} - u_{n-4} u_{n-2} + u_{n-4} u_{n-1}) ]
\nonumber \\
&& + u_n 
[(u_{n+1} + u_{n+2} + u_{n+3})^2 - u_n (u_{n-3} + u_{n-2} + u_{n-1})^2] 
\nonumber \\
&& + u_n^2 [ u_{n+1} + u_{n+2} + u_{n+3} - (u_{n-3} + u_{n-2} + u_{n-1})].
\end{eqnarray}


\begin{thebibliography}{99}

\bibitem{ghjsc} 
\"{U}. G\"{o}kta\c{s} and W. Hereman,
{\em Symbolic computation of conserved densities for systems of nonlinear
evolution equations}, J. Symbolic Computation, 24 (1997), pp.~591--621.

\bibitem{physicad} 
\sameauthor,
{\em Computation of conservation laws for nonlinear lattices}, 
Physica D (1998) to appear.

\bibitem{baltzer}
\sameauthor, 
{\em Computation of higher-order symmetries for 
nonlinear evolution and lattice equations}, 
Adv. in Comp. Math. (1998) submitted.

\bibitem{soft}
\sameauthor, 
The {\em Mathematica} Package {\bf InvariantsSymmetries.m} and the related 
files are available at 
http://www.mathsource.com/cgi-bin/msitem?0208-932.
{\it MathSource\/} is an electronic library of {\it Mathematica\/} 
material maintained by Wolfram Research, Inc. (Champaign, Illinois, U.S.A.).

\bibitem{ghepla}
\"{U}. G\"{o}kta\c{s}, W. Hereman and G. Erdmann,
{\em Computation of conserved densities for systems of nonlinear
differential-difference equations},  Phys. Lett. A, 236 (1997) pp.~30--38. 

\bibitem{itoh}
Y. Itoh, 
{\em Integrals of a Lotka-Volterra system of odd number of variables},
Prog. Theor. Phys., 78 (1987) pp. ~507--510.

\bibitem{verheest}
F. Verheest, 
{\em Integrability, invariants and bi-Hamiltonian structure of vector
nonlinear evolution equations}, in these Proceedings (1998).

\end{thebibliography}
\end{document}